# Studies on properties and estimation problems for modified extension of exponential distribution

**M. A. El-Damcese[1], Dina. A. Ramadan[2]**

[1] Mathematics Department, Faculty of Science, Tanta University, Tanta, Egypt

(e-mail:meldamcese@yahoo.com)

[2] Mathematics Department, Faculty of Science, Mansoura University, Mansoura, Egypt

(e-mail:Dina_Ahmed2188@yahoo.com)

**Abstract**: The present paper considers modified extension of the exponential distribution with three parameters. We study the main properties of this new distribution, with special emphasis on its median, mode and moments function and some characteristics related to reliability studies. For Modified- extension exponential distribution (MEXED) we have obtained the Bayes Estimators of scale and shape parameters using Lindley's approximation (L-approximation) under squared error loss function. But, through this approximation technique it is not possible to compute the interval estimates of the parameters. Therefore, we also propose Gibbs sampling method to generate sample from the posterior distribution. On the basis of generated posterior sample we computed the Bayes estimates of the unknown parameters and constructed 95 % highest posterior density credible intervals. A Monte Carlo simulation study is carried out to compare the performance of Bayes estimators with the corresponding classical estimators in terms of their simulated risk. A real data set has been considered for illustrative purpose of the study.

## 1. Introduction

In the field of lifetime modelling exponential distribution (ED) has greater importance to study the reliability characteristics of any lifetime phenomenon. The popularity of this model has been discussed by several authors. Although it become most popular due to its constant failure rate pattern, but in many practical situation this distribution is not suited to study the phenomenon where failure rate is not constant. In recent years, several new classes of models were introduced based on modification of exponential distribution. For example, *Gupta* and *Kundu* (1999) and *Gupta* and *Kundu* (2001) introduced an extension of the exponential distribution typically called the generalized exponential (GE) distribution. Therefore, it is said that the random variable *x* follows the GE distribution if its density function is given by

$$g_1(x;\alpha,\beta) = \alpha\beta\, e^{-\alpha x}(1-e^{-\alpha x})^{\beta-1},$$

(1)

where $x > 0, \alpha > 0$ and $\beta > 0$. We use the notation $X \sim GE(\alpha,\beta)$ for a random variable with such distribution.

More recently, *Nadarajah* and *Haghighi* (2011) introduced another extension of the exponential model, so that a random variable X follows the Nadarajah and Haghighi's exponential distribution (NHE) if its density function is given by

$$g_2(x;\alpha,\beta) = \alpha\beta(1+\alpha x^2)^{\beta-1} e^{\left[1-(1+\alpha x^2)^\beta\right]},$$

(2)

where $x > 0, \alpha > 0$ and $\beta > 0$. We use the notation $X \sim NHE(\alpha,\beta)$. *Sanjay et al*. (2014) explained the classical and Bayesian estimation of unknown parameters and reliability characteristics in extension of exponential distribution.

Both distributions have the exponential distribution (E) with scale parameter $\alpha$, as a special case when $\beta = 1$, that is,

$$g_1(x;\alpha,\beta=1) = g_2(x;\alpha,\beta=1) = \alpha\, e^{-\alpha x},$$

(3)

where $x > 0$ and $\alpha > 0$ with the notation $X \sim E(\alpha)$. Other extensions of the exponential model in the survival analysis context are considered in the *Marshall* and *Olkin*'s (2007) book.

The main object of this paper is to present yet another extension for the exponential distribution that can be used as an alternative to the ones mentioned above. We discuss some properties for this new distribution. We consider the classical and Bayesian estimation of the unknown parameters and reliability characteristics of a new extension of exponential distribution. It is observed that the MLEs of the unknown parameters can not be obtained in nice closed form, as expected, and they have to obtain by solving two nonlinear equations simultaneously. It is remarkable that most of the Bayesian inference procedures have been developed with the usual squared-error loss function, which is symmetrical and associates equal importance to the losses due to overestimation and underestimation of equal magnitude. However, such a restriction may be impractical in most situations of practical importance. For example, in the estimation of reliability and failure rate functions, an overestimation is usually much more serious than an underestimation. In this case, the use of symmetrical loss function might be inappropriate as also emphasized by **Basu** and **Ebrahimi** (1991). Further, we consider the Bayesian inference of the unknown parameters under the assumption that both parameters have independent gamma priors. It is observed that the Bayes estimators have not been obtained in explicit form. Therefore, Lindley's approximation method is used. Unfortunately, by using Lindley's approximation method it is not possible to construct the highest posterior density (HPD) credible intervals. Therefore, we have also used Monte Carlo Markov Chain method (Gibbs sampling procedure) to construct the 95% HPD credible intervals for the parameters and estimates are also coded on the basis of MCMC samples. Monte Carlo simulations are conducted to compare the performances of the classical estimators with corresponding Bayes estimators obtained under squared error loss function in both informative and non-informative set-up for complete sample. Further, we have also constructed 95% approximate confidence intervals and highest posterior density (HPD) credible intervals for the parameters.

## 2. Density and Properties

A random variable X is distributed according to the modified extended exponential distribution (MExED) with parameters $\alpha, \lambda$ and $\beta$ if its density function and the cumulative distribution function of this new family of distribution can be given as

$$f(x) = \alpha(\lambda + 2\beta x)(1 + \lambda x + \beta x^2)^{\alpha-1} e^{\left[1-(1+\lambda x+\beta x^2)^{\alpha}\right]}$$

(4)

where $x \geq 0, \alpha \geq 0, \lambda \geq 0$ and $\beta \geq 0$. We use the notation $X \sim MExED(\alpha, \lambda, \beta)$.

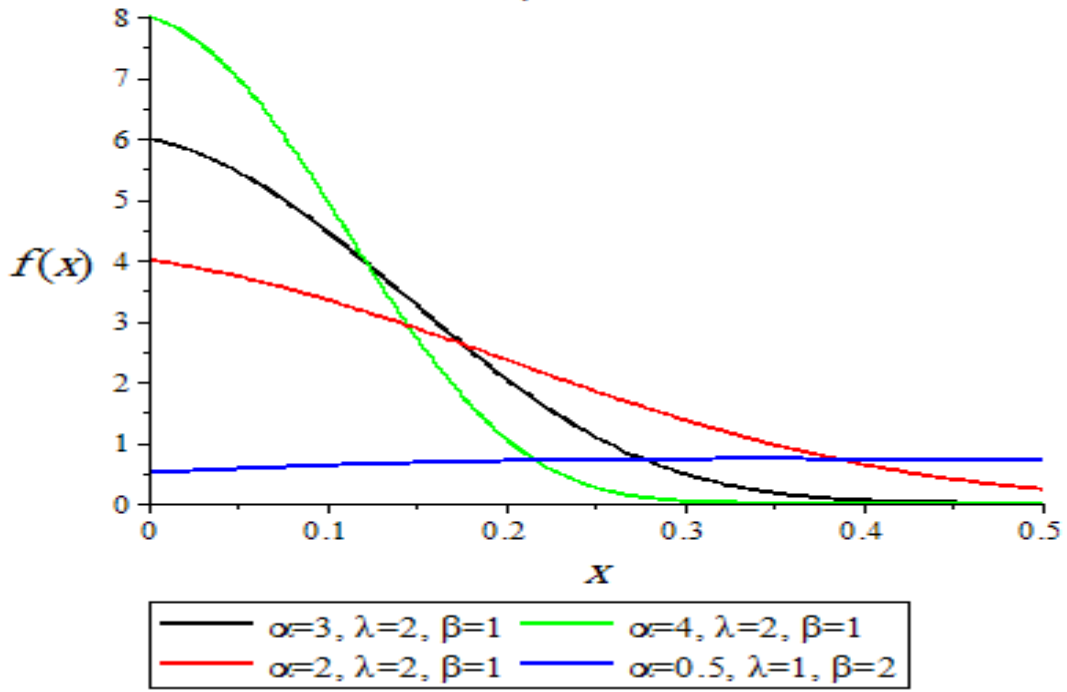

Figure1: Density plot for various choice for α, λ and β

and

$$F(x) = 1 - e^{\left[1-(1+\lambda x+\beta x^2)^\alpha\right]}$$

(5)

where $x \geq 0, \alpha \geq 0, \lambda \geq 0$ and $\beta \geq 0$.

The modified extended exponential distribution (MExED) can be a useful characterization of life time data analysis. The reliability function (R) of the modified extended exponential distribution (MExED) is denoted by $R(t)$ also known as the survivor function and is defined as

$$R(t) = e^{\left[1-(1+\lambda t+\beta t^2)^\alpha\right]}; \quad t, \alpha, \lambda, \beta \geq 0$$

(6)

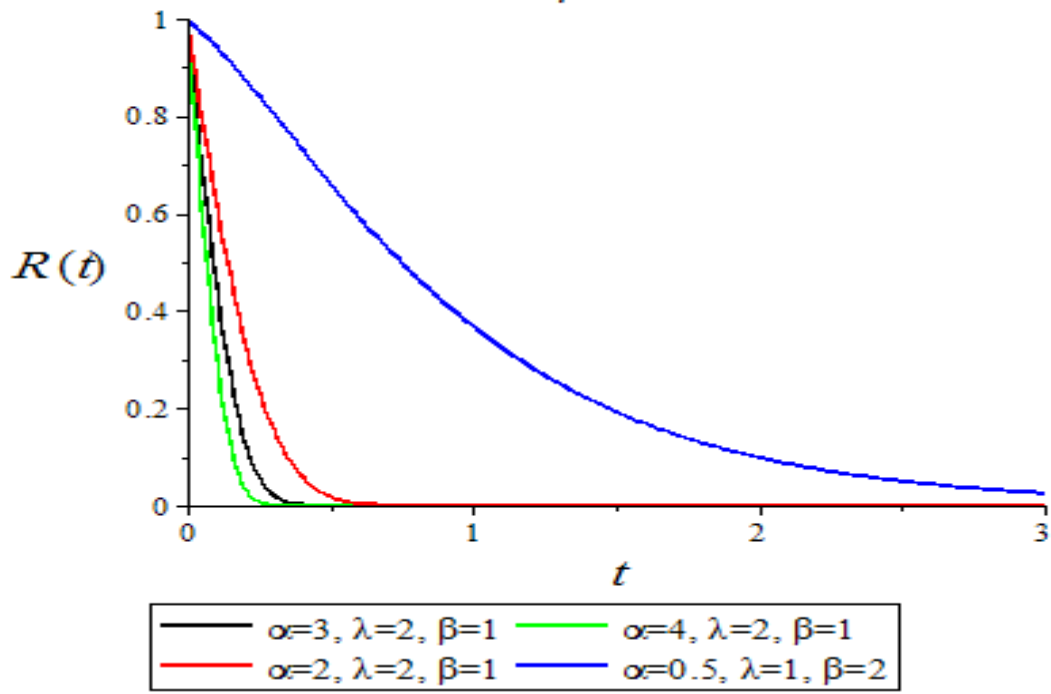

**Figure 2:** Reliability plot for various choice for α, λ and β

One of the characteristic in reliability analysis is the hazard rate function (*HRF*) defined by

$$h(t) = \frac{f(t)}{R(t)} = \alpha(\lambda + 2\beta t)(1 + \lambda t + \beta t^2)^{\alpha-1}; \quad t, \alpha, \lambda, \beta \geq 0.$$

(7)

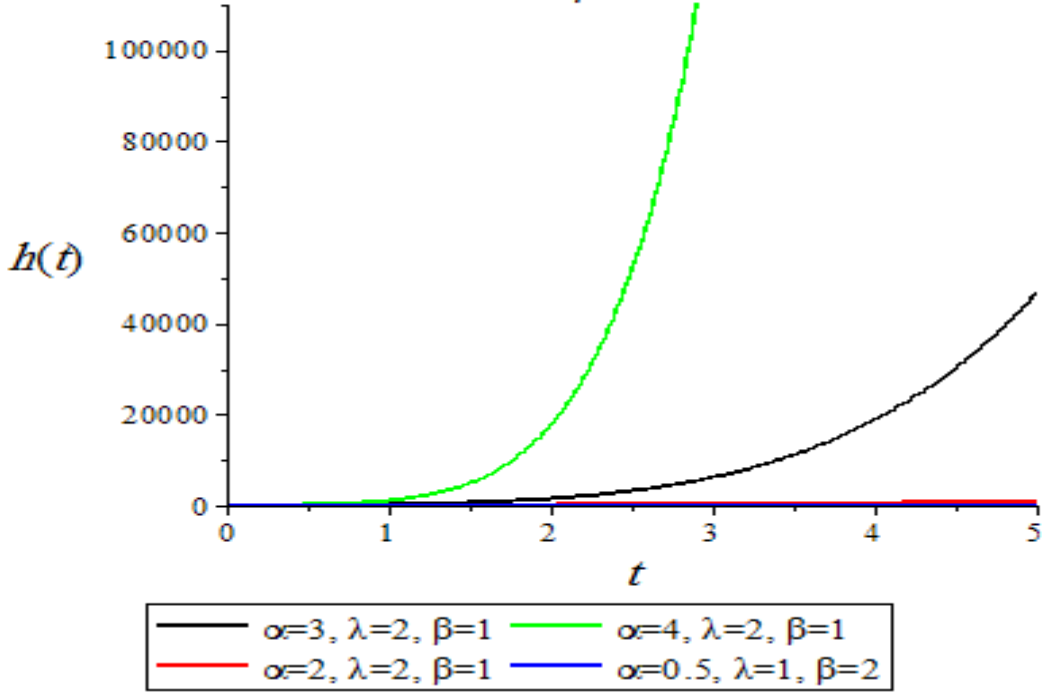

**Figure 3:** Failure Rate plot for various choice for α, λ and β

It is important to note that the units for $h(t)$ is the probability of failure per unit of time, distance or cycles. These failure rates are defined with different choices of parameters. The cumulative hazard function of the modified extended exponential distribution is denoted by $H(t)$ and is defined as

$$H(t) = \int_0^t h(x)dt = \int_0^t \alpha(\lambda + 2\beta x)(1 + \lambda x + \beta x^2)^{\alpha-1} dx$$
$$= (1 + \lambda t + \beta t^2)^\alpha - 1.$$

(8)

## 3. Statistical Analysis

### 3.1 The Median and Mode

It is observed as expected that the mean of MExED($\alpha, \lambda, \beta$) cannot be obtained in explicit forms. It can be obtained as infinite series expansion so, in general different moments of MExED($\alpha, \lambda, \beta$). Also, we cannot get the quantile $x_q$ of MExED($\alpha, \lambda, \beta$) in a closed form by using the equation $F_X(x_q; \alpha, \lambda, \beta) - q = 0$. Thus, by using Equation (5), we find that

$$(\lambda x_q + \beta x_q^2) = [1 - \ln(1-q)]^{1/\alpha} - 1, 0 < q < 1.$$

(9)

The median $m(X)$ of MExED$(\alpha, \lambda, \beta)$ can be obtained from (9), when $q = 0.5$, as follows

$$(\lambda x_{0.5} + \beta x_{0.5}^2) = [1 - \ln(0.5)]^{1/\alpha} - 1.$$

(10)

Moreover, the mode of MExED$(\alpha, \lambda, \beta)$ can be obtained as a solution of the following nonlinear equation.

$$\frac{d}{dx} f_X(x; \alpha, \lambda, \beta) = 0,$$

$$\frac{d}{dx}\left[\alpha(\lambda + 2\beta x)(1 + \lambda x + \beta x^2)^{\alpha-1} e^{\left[1-(1+\lambda x+\beta x^2)^\alpha\right]}\right] = 0.$$

(11)

### 3.2 Moment

The $r^{th}$ moments of the MExED is denoted by $\mu_r'$ and it is given by

$$\mu_r' = \sum_{n=m=0}^{\infty} \binom{r}{n}\binom{\frac{r-n}{2}}{m}(-1)^n e^1 2^{2m-r} \lambda^n \beta^{m-r}(\lambda^2 - 4\beta)^{\frac{r-n-2m}{2}} \Gamma\left(\frac{m}{\alpha}+1, 1\right),$$

(12)

The mean and variance of MExED are

$$E(x) = \sum_{n=m=0}^{\infty} \binom{1}{n}\binom{\frac{1-n}{2}}{m}(-1)^n e^1 2^{2m-1} \lambda^n \beta^{m-1}(\lambda^2 - 4\beta)^{\frac{-n-2m+1}{2}} \Gamma\left(\frac{m}{\alpha}+1, 1\right)$$

(13)

and

$$Var(x) = \sum_{n=m=0}^{\infty} \binom{2}{n}\binom{\frac{2-n}{2}}{m}(-1)^n e^1 4^{m-1} \lambda^n \beta^{m-2}(\lambda^2 - 4\beta)^{\frac{-n-2m+2}{2}} \Gamma\left(\frac{m}{\alpha}+1, 1\right)$$

$$- \left[\sum_{n=m=0}^{\infty} \binom{1}{n}\binom{\frac{1-n}{2}}{m}(-1)^n e^1 2^{2m-1} \lambda^n \beta^{m-1}(\lambda^2 - 4\beta)^{\frac{-n-2m+1}{2}} \Gamma\left(\frac{m}{\alpha}+1, 1\right)\right]^2.$$

(14)

### 4. Classical estimation

In this section, we have obtained the maximum likelihood estimates (MLEs) of the parameters, reliability function and hazard function for the considered model. Let us suppose that n units are put on a test with corresponding life times being identically distributed with probability density function (4) and cumulative distribution function

(5). Then, the likelihood function can be written as

$$L(x\backslash\alpha,\lambda,\beta) = \prod_{i=1}^{n} f(x_i; \alpha, \lambda, \beta)$$

$$= \alpha^n e^{\sum_{i=1}^{n}(1-(1+\lambda x_i+\beta x_i^2)^\alpha)} \prod_{i=1}^{n}(\lambda + 2\beta x_i)(1+\lambda x_i + \beta x_i^2)^{\alpha-1},$$

(15)

$$\ln L(\alpha, \lambda, \beta) = n \ln \alpha + \sum_{i=1}^{n} \ln(\lambda + 2\beta x_i) + (\alpha - 1)\sum_{i=1}^{n}\ln(1+\lambda x_i + \beta x_i^2)$$

$$+ \sum_{i=1}^{n}(1-(1+\lambda x_i + \beta x_i^2)^\alpha),$$

(16)

$$\frac{\partial \ln L}{\partial \lambda} = \sum_{i=1}^{n}\frac{1}{\lambda + 2\beta x_i} + (\alpha - 1)\sum_{i=1}^{n}\frac{x_i}{1+\lambda x_i + \beta x_i^2} - \sum_{i=1}^{n}\alpha x_i(1+\lambda x_i + \beta x_i^2)^{\alpha-1},$$

(17)

$$\frac{\partial \ln L}{\partial \beta} = \sum_{i=1}^{n}\frac{2x_i}{\lambda + 2\beta x_i} + (\alpha - 1)\sum_{i=1}^{n}\frac{x_i^2}{1+\lambda x_i + \beta x_i^2} - \sum_{i=1}^{n}\alpha x_i^2(1+\lambda x_i + \beta x_i^2)^{\alpha-1},$$

(18)

$$\frac{\partial \ln L}{\partial \alpha} = \frac{n}{\alpha} + \sum_{i=1}^{n}\ln(1+\lambda x_i + \beta x_i^2) - \sum_{i=1}^{n}(1+\lambda x_i + \beta x_i^2)^\alpha \ln(1+\lambda x_i + \beta x_i^2),$$

(19)

Maximum likelihood estimates can be obtained by solving the above two equations simultaneously, but these equations cannot be expressed in explicit form. Therefore, Non linear maximization technique (in built command in R software) has been used to compute the MLEs of the parameters. Further, let $(\hat{\alpha}, \hat{\lambda}, \hat{\beta})$ are the MLEs of α, λ and β respectively. Therefore, using invariance property of MLEs, the Bayes estimators of reliability function $\hat{R}$ and hazard function $\hat{h}$ for any specified time t are given by following equations.

$$\hat{R}(t) = e^{\left[1-(1+\hat{\lambda}t+\hat{\beta}t^2)^{\hat{\alpha}}\right]}$$

(20)

and

$$\hat{h}(t) = \hat{\alpha}(\hat{\lambda} + 2\hat{\beta}t)(1+\hat{\lambda}t + \hat{\beta}t^2)^{\hat{\alpha}-1}.$$

## 4.1 Asymptotic Intervals for the Parameters

In this subsection, we obtained the Fisher information matrix to compute 95% asymptotic confidence intervals for the parameters based on maximum likelihood estimators (MLEs). The Fisher information matrix can be obtained by using log-likelihood function (16). Thus we have

$$I(\hat{\alpha}, \hat{\lambda}, \hat{\beta}) = \begin{pmatrix} -\dfrac{\partial^2 \ln L}{\partial \lambda^2} & -\dfrac{\partial^2 \ln L}{\partial \lambda \partial \beta} & -\dfrac{\partial^2 \ln L}{\partial \lambda \partial \alpha} \\ -\dfrac{\partial^2 \ln L}{\partial \beta \partial \lambda} & -\dfrac{\partial^2 \ln L}{\partial \beta^2} & -\dfrac{\partial^2 \ln L}{\partial \beta \partial \alpha} \\ -\dfrac{\partial^2 \ln L}{\partial \alpha \partial \lambda} & -\dfrac{\partial^2 \ln L}{\partial \alpha \partial \beta} & -\dfrac{\partial^2 \ln L}{\partial \alpha^2} \end{pmatrix}$$

(22)

where,

$$\frac{\partial^2 \ln L}{\partial \lambda^2} = \sum_{i=1}^{n} \frac{-1}{(\lambda + 2\beta x_i)^2} - (\alpha - 1) \sum_{i=1}^{n} \frac{x_i^2}{(1 + \lambda x_i + \beta x_i^2)^2} - \sum_{i=1}^{n} (\alpha^2 - \alpha) x_i^2 (1 + \lambda x_i + \beta x_i^2)^{\alpha - 2},$$

(23)

$$\frac{\partial^2 \ln L}{\partial \lambda \partial \beta} = \sum_{i=1}^{n} \frac{-2 x_i}{(\lambda + 2\beta x_i)^2} - (\alpha - 1) \sum_{i=1}^{n} \frac{x_i^3}{(1 + \lambda x_i + \beta x_i^2)^2} - \sum_{i=1}^{n} (\alpha^2 - \alpha) x_i^3 (1 + \lambda x_i + \beta x_i^2)^{\alpha - 2},$$

(24)

$$\frac{\partial^2 \ln L}{\partial \lambda \partial \alpha} = \sum_{i=1}^{n} \frac{x_i}{1 + \lambda x_i + \beta x_i^2} - \sum_{i=1}^{n} x_i (1 + \lambda x_i + \beta x_i^2)^{\alpha - 1} [1 + \alpha \ln(1 + \lambda x_i + \beta x_i^2)],$$

(25)

$$\frac{\partial^2 \ln L}{\partial \beta^2} = \sum_{i=1}^{n} \frac{-4 x_i^2}{(\lambda + 2\beta x_i)^2} - (\alpha - 1) \sum_{i=1}^{n} \frac{x_i^4}{(1 + \lambda x_i + \beta x_i^2)^2} - \sum_{i=1}^{n} (\alpha^2 - \alpha) x_i^4 (1 + \lambda x_i + \beta x_i^2)^{\alpha - 2},$$

(26)

$$\frac{\partial^2 \ln L}{\partial \beta \partial \alpha} = \sum_{i=1}^{n} \frac{x_i^2}{1 + \lambda x_i + \beta x_i^2}$$

$$- \sum_{i=1}^{n} x_i^2 (1 + \lambda x_i + \beta x_i^2)^{\alpha-1}[1 + \alpha \ln(1 + \lambda x_i + \beta x_i^2)],$$

(27)

$$\frac{\partial^2 \ln L}{\partial \alpha^2} = \frac{-n}{\alpha^2} - \sum_{i=1}^{n} (1 + \lambda x_i + \beta x_i^2)^{\alpha}[\ln(1 + \lambda x_i + \beta x_i^2)]^2.$$

(28)

All the above derivatives are evaluated at $(\hat{\alpha}, \hat{\lambda}, \hat{\beta})$. The above matrix can be inverted to obtain the estimate of the asymptotic variance-covariance matrix of the MLEs and diagonal elements of $I^{-1}(\hat{\alpha}, \hat{\lambda}, \hat{\beta})$ provides asymptotic variance of $\alpha, \lambda$ and $\beta$ respectively. The above approach is used to derive the $100(1 - \gamma)\%$ confidence intervals of the parameters $\alpha, \lambda, \beta$ as in the following forms

$$\hat{\alpha} \pm Z_{\gamma/2}\sqrt{Var(\hat{\alpha})}, \quad \hat{\lambda} \pm Z_{\gamma/2}\sqrt{Var(\hat{\lambda})} \text{ and } \hat{\beta} \pm Z_{\gamma/2}\sqrt{Var(\hat{\beta})}.$$

(29)

## 5. Bayesian Estimation of the Parameters

In this section, we have derived the expression posterior distributions for the considered model. Let $X = (x_1, x_2, x_3, \ldots, x_n)$ be a random sample of size n observed from (4), and then the likelihood function is given as in (15). As we seen that this model is a good alternative of the several exponentiated family and reduces in exponential family for a $\alpha = 1$ and $\beta = 0$. Since for this distribution not a single conjugate prior is known till date. Therefore, we consider independent gamma priors for shape i.e. $\alpha \sim gamma(a, b)$ as well as scale parameter i.e $\lambda \sim gamma(c, d)$ and $\beta \sim gamma(g, f)$. Therefore, the joint prior of $(\alpha, \lambda, \beta)$ is given as

$$\pi(\alpha, \lambda, \beta) \propto \alpha^{a-1} \lambda^{c-1} \beta^{g-1} e^{-b\alpha - d\lambda - f\beta}$$

(30)

where a,b,c,d,g and f are the hyper parameters. Therefore, the joint posterior distribution can written as,

$$P(\alpha, \lambda, \beta | X) \propto \alpha^{n+a-1} \lambda^{n+c-1} \beta^{g-1} e^{-b\alpha - d\lambda - f\beta} e^{\sum_{i=1}^{n}\left(1-(1+\lambda x_i + \beta x_i^2)^\alpha\right)} \prod_{i=1}^{n}\left(1 + \frac{2\beta}{\lambda}x_i\right)(1+\lambda x_i + \beta x_i^2)^{\alpha-1}.$$

(31)

Under squared error loss function (SELF) the Bayes estimate is the posterior mean of the distribution. Therefore, the Bayes estimate of $(\alpha, \lambda, \beta)$, Reliability function $R(t)$ and Hazard function $h(t)$ can be expressed in following equations.

$$\hat{\alpha} = K^{-1} \int_0^\infty \int_0^\infty \int_0^\infty \alpha^{n+a} \lambda^{n+c-1} \beta^{g-1} e^{-b\alpha - d\lambda - f\beta} e^{\sum_{i=1}^{n}\left(1-(1+\lambda x_i + \beta x_i^2)^\alpha\right)} \prod_{i=1}^{n}\left(1 + \frac{2\beta}{\lambda}x_i\right)(1+\lambda x_i + \beta x_i^2)^{\alpha-1} d\beta\, d\lambda\, d\alpha,$$

(32)

$$\hat{\lambda} = K^{-1} \int_0^\infty \int_0^\infty \int_0^\infty \alpha^{n+a-1} \lambda^{n+c} \beta^{g-1} e^{-b\alpha - d\lambda - f\beta} e^{\sum_{i=1}^{n}\left(1-(1+\lambda x_i + \beta x_i^2)^\alpha\right)} \prod_{i=1}^{n}\left(1 + \frac{2\beta}{\lambda}x_i\right)(1+\lambda x_i + \beta x_i^2)^{\alpha-1} d\beta\, d\lambda\, d\alpha,$$

(33)

$$\hat{\beta} = K^{-1} \int_0^\infty \int_0^\infty \int_0^\infty \alpha^{n+a-1} \lambda^{n+c-1} \beta^{g} e^{-b\alpha - d\lambda - f\beta} e^{\sum_{i=1}^{n}\left(1-(1+\lambda x_i + \beta x_i^2)^\alpha\right)} \prod_{i=1}^{n}\left(1 + \frac{2\beta}{\lambda}x_i\right)(1+\lambda x_i + \beta x_i^2)^{\alpha-1} d\beta\, d\lambda\, d\alpha,$$

(34)

$$\hat{R}(t) = K^{-1} \int_0^\infty \int_0^\infty \int_0^\infty \alpha^{n+a-1} \lambda^{n+c-1} \beta^{g-1} e^{\left(1-b\alpha - d\lambda - f\beta - (1+\lambda t + \beta t^2)^\alpha\right)}$$
$$e^{\sum_{i=1}^{n}\left(1-(1+\lambda x_i + \beta x_i^2)^\alpha\right)} \prod_{i=1}^{n}\left(1 + \frac{2\beta}{\lambda}x_i\right)(1+\lambda x_i + \beta x_i^2)^{\alpha-1} d\beta\, d\lambda\, d\alpha$$

(35)

and

$$\hat{h}(t) = K^{-1} \int_0^\infty \int_0^\infty \int_0^\infty \alpha^{n+a} \lambda^{n+c} \beta^{g-1} \left(1 + \frac{2\beta}{\lambda}t\right)(1+\lambda t + \beta t^2)^{\alpha-1} e^{-b\alpha - d\lambda - f\beta}$$
$$e^{\sum_{i=1}^{n}\left(1-(1+\lambda x_i + \beta x_i^2)^\alpha\right)} \prod_{i=1}^{n}\left(1 + \frac{2\beta}{\lambda}x_i\right)(1+\lambda x_i + \beta x_i^2)^{\alpha-1} d\beta\, d\lambda\, d\alpha.$$

(36)

where

$$K = \int_0^\infty \int_0^\infty \int_0^\infty \alpha^{n+a-1} \lambda^{n+c-1} \beta^{g-1} e^{-b\alpha - d\lambda - f\beta} e^{\sum_{i=1}^n \left(1 - (1+\lambda x_i + \beta x_i^2)^\alpha\right)} \prod_{i=1}^n \left(1 + \frac{2\beta}{\lambda} x_i\right) (1 + \lambda x_i + \beta x_i^2)^{\alpha-1} \, d\beta \, d\lambda \, d\alpha.$$

(37)

From the above, it is easy to observed that the analytical solution of the Bayes estimators are not possible. Therefore, we have used the Lindley's approximation methods and Markov Chain Monte Carlo method to obtain the approximate solutions of the above Eqs. (32– 36).

## 5.1 Lindley's Approximation

It may be noted here that the posterior distribution of $(\alpha, \lambda, \beta)$ takes a ratio form that involves an integration in the denominator and cannot be reduced to a closed form. Hence, the evaluation of the posterior expectation for obtaining the Bayes estimator of α, λ and β will be tedious. Among the various methods suggested to approximate the ratio of integrals of the above form, perhaps the simplest one is *Lindley's* (1980) approximation method, which approaches the ratio of the integrals as a whole and produces a single numerical result. Many authors have used this approximation for obtaining the Bayes estimators for some lifetime distributions; see among others, ***Howlader*** and ***Hossain*** (2002) and ***Jaheen*** (2005).

Thus, we propose the use of ***Lindley's*** (1980) approximation for obtaining the Bayes estimator of α, λ and β by considering the function $I(x)$, defined as follows;

$$I(x) = E[u(\alpha, \lambda, \beta)] = \frac{\int u(\alpha, \lambda, \beta) e^{L(\alpha,\lambda,\beta)+G(\alpha,\lambda,\beta)} \, d(\alpha, \lambda, \beta)}{\int e^{L(\alpha,\lambda,\beta)+G(\alpha,\lambda,\beta)} \, d(\alpha, \lambda, \beta)},$$

(38)

where

$u(\alpha, \lambda, \beta)$ is a function of $\alpha, \lambda$ and $\beta$ only

$L(\alpha, \lambda, \beta)$ is log of likelihood

$G(\alpha, \lambda, \beta)$ is log joint prior of $\alpha, \lambda$ and $\beta$,

According to ***Lindley*** (1980), if ML estimates of the parameters are available and n is sufficiently large then the above ratio of the integral can be approximated as:

$$I(x) = u(\hat{\alpha}, \hat{\lambda}, \hat{\beta}) + (u_1 a_1 + u_2 a_2 + u_3 a_3 + a_4 + a_5)$$
$$+ \frac{1}{2}[A\,(u_1\sigma_{11}+u_2\sigma_{12}+u_3\sigma_{13})+B(u_1\sigma_{21}+u_2\sigma_{22}+u_3\sigma_{23})+C(u_1\sigma_{31}+u_2\sigma_{32}+u_3\sigma_{33})]$$

(39)

where

$a_i = \rho_1\sigma_{i1}+\rho_2\sigma_{i2}+\rho_3\sigma_{i3}, \quad i = 1,2,3$

$a_4 = u_{12}\sigma_{12} + u_{13}\sigma_{13} + u_{23}\sigma_{23}$

$a_5 = \frac{1}{2}(u_{11}\sigma_{11} + u_{22}\sigma_{22} + u_{33}\sigma_{33})$

$A = \sigma_{11}L_{111} + 2\sigma_{12}L_{121} + 2\sigma_{13}L_{131}+2\sigma_{23}L_{231} + \sigma_{22}L_{221} + \sigma_{33}L_{331}$

$B = \sigma_{11}L_{112} + 2\sigma_{12}L_{122} + 2\sigma_{13}L_{132}+2\sigma_{23}L_{232} + \sigma_{22}L_{222} + \sigma_{33}L_{332}$

$C = \sigma_{11}L_{113} + 2\sigma_{12}L_{123} + 2\sigma_{13}L_{133}+2\sigma_{23}L_{233} + \sigma_{22}L_{223} + \sigma_{33}L_{333}$

and subscripts 1, 2, 3 on the right-hand sides refer to $\alpha, \lambda, \beta$ respectively and let $\theta_1 = \alpha, \theta_2 = \beta$ and $\theta_3 = \lambda$

$$\rho_i = \frac{\partial \rho}{\partial \theta_i}, u_i = \frac{\partial u(\theta_1, \theta_2, \theta_3)}{\partial \theta_i}, i = 1,2,3,$$

$$u_{ij} = \frac{\partial^2 u(\theta_1, \theta_2, \theta_3)}{\partial \theta_i \, \partial \theta_j}, i,j = 1,2,3,$$

$$L_{ij} = \frac{\partial^2 L(\theta_1, \theta_2, \theta_3)}{\partial \theta_i \, \partial \theta_j}, i,j = 1,2,3,$$

$$L_{ijk} = \frac{\partial^3 L(\theta_1, \theta_2, \theta_3)}{\partial \theta_i \, \partial \theta_j \, \partial \theta_k}, i,j,k = 1,2,3.$$

and $\sigma_{ij}$ is the $(i,j)$−th element of the inverse of the matrix $\{L_{ij}\}$, all evaluated at the MLE of parameters.

For the prior distribution (30) we have

$$\rho = \ln \pi(\alpha, \lambda, \beta) = (a-1)\ln\alpha + (c-1)\ln\lambda + (g-1)\ln\beta - (b\alpha + d\lambda + f\beta)$$

and then we get

$$\rho_1 = \frac{a-1}{\alpha} - b, \rho_2 = \frac{c-1}{\lambda} - d, \rho_3 = \frac{g-1}{\beta} - f$$

Also, the values of $L_{ij}$ can be obtained as follows for $i,j = 1,2,3$

$$L_{11} = \frac{-n}{\alpha^2} - \sum_{i=1}^{n}(1+\lambda x_i + \beta x_i^2)^\alpha [\ln(1+\lambda x_i + \beta x_i^2)]^2,$$

$$L_{12} = L_{21} = \sum_{i=1}^{n} \frac{x_i}{1 + \lambda x_i + \beta x_i^2}$$

$$- \sum_{i=1}^{n} x_i(1 + \lambda x_i + \beta x_i^2)^{\alpha-1}[1 + \alpha \ln(1 + \lambda x_i + \beta x_i^2)],$$

$$L_{13} = L_{31} = \sum_{i=1}^{n} \frac{x_i^2}{1 + \lambda x_i + \beta x_i^2}$$

$$- \sum_{i=1}^{n} x_i^2(1 + \lambda x_i + \beta x_i^2)^{\alpha-1}[1 + \alpha \ln(1 + \lambda x_i + \beta x_i^2)],$$

$$L_{22} = \sum_{i=1}^{n} \frac{-1}{(\lambda + 2\beta x_i)^2} - (\alpha - 1)\sum_{i=1}^{n} \frac{x_i^2}{(1 + \lambda x_i + \beta x_i^2)^2}$$

$$- \sum_{i=1}^{n} (\alpha^2 - \alpha)x_i^2(1 + \lambda x_i + \beta x_i^2)^{\alpha-2},$$

$$L_{23} = L_{32} = \sum_{i=1}^{n} \frac{-2x_i}{(\lambda + 2\beta x_i)^2} - (\alpha - 1)\sum_{i=1}^{n} \frac{x_i^3}{(1 + \lambda x_i + \beta x_i^2)^2}$$

$$- \sum_{i=1}^{n} (\alpha^2 - \alpha)x_i^3(1 + \lambda x_i + \beta x_i^2)^{\alpha-2},$$

$$L_{33} = \sum_{i=1}^{n} \frac{-4x_i^2}{(\lambda + 2\beta x_i)^2} - (\alpha - 1)\sum_{i=1}^{n} \frac{x_i^4}{(1 + \lambda x_i + \beta x_i^2)^2}$$

$$- \sum_{i=1}^{n} (\alpha^2 - \alpha)x_i^4(1 + \lambda x_i + \beta x_i^2)^{\alpha-2}.$$

and the values of $L_{ijk}$ for $i, j, k = 1,2,3$

$$L_{111} = \frac{2n}{\alpha^3} - \sum_{i=1}^{n} (1 + \lambda x_i + \beta x_i^2)^{\alpha} \ln(1 + \lambda x_i + \beta x_i^2)^3,$$

$L_{112} == L_{121} = L_{211}$

$$= -\sum_{i=1}^{n} x_i(1 + \lambda x_i + \beta x_i^2)^{\alpha-1} \ln(1 + \lambda x_i + \beta x_i^2) [\alpha \ln(1 + \lambda x_i + \beta x_i^2) + 2],$$

$L_{113} = L_{131} = L_{311}$

$$= -\sum_{i=1}^{n} x_i^2 (1 + \lambda x_i + \beta x_i^2)^{\alpha-1} \ln(1 + \lambda x_i + \beta x_i^2) [\alpha \ln(1 + \lambda x_i + \beta x_i^2) + 2],$$

$L_{122} = L_{212} = L_{221}$

$$= \sum_{i=1}^{n} \frac{-x_i^2}{(1 + \lambda x_i + \beta x_i^2)^2}$$

$$- \sum_{i=1}^{n} x_i^2 (1 + \lambda x_i + \beta x_i^2)^{\alpha-2} [(2\alpha - 1) + \alpha(\alpha - 1) \ln(1 + \lambda x_i + \beta x_i^2)],$$

$L_{133} = L_{313} = L_{331}$

$$= \sum_{i=1}^{n} \frac{-x_i^4}{(1 + \lambda x_i + \beta x_i^2)^2}$$

$$- \sum_{i=1}^{n} x_i^4 (1 + \lambda x_i + \beta x_i^2)^{\alpha-2} [(2\alpha - 1) + \alpha(\alpha - 1) \ln(1 + \lambda x_i + \beta x_i^2)],$$

$L_{123} = L_{213} = L_{132} = L_{312} = L_{231} = L_{321}$

$$= \sum_{i=1}^{n} \frac{-x_i^3}{(1 + \lambda x_i + \beta x_i^2)^2}$$

$$- \sum_{i=1}^{n} x_i^3 (1 + \lambda x_i + \beta x_i^2)^{\alpha-2} [(2\alpha - 1) + \alpha(\alpha - 1) \ln(1 + \lambda x_i + \beta x_i^2)],$$

$$L_{222} = \sum_{i=1}^{n} \frac{2}{(\lambda + 2\beta x_i)^3} + (\alpha - 1) \sum_{i=1}^{n} \frac{2x_i^3}{(1 + \lambda x_i + \beta x_i^2)^3}$$

$$- \sum_{i=1}^{n} \alpha(\alpha - 1)(\alpha - 2) x_i^3 (1 + \lambda x_i + \beta x_i^2)^{\alpha-3},$$

$L_{223} = L_{232} = L_{322} =$

$$= \sum_{i=1}^{n} \frac{4x_i}{(\lambda + 2\beta x_i)^3} + (\alpha - 1) \sum_{i=1}^{n} \frac{2x_i^4}{(1 + \lambda x_i + \beta x_i^2)^3}$$

$$- \sum_{i=1}^{n} \alpha(\alpha - 1)(\alpha - 2) x_i^4 (1 + \lambda x_i + \beta x_i^2)^{\alpha-3},$$

$$L_{233} = L_{323} = L_{332} =$$

$$= \sum_{i=1}^{n} \frac{8x_i^2}{(\lambda + 2\beta x_i)^3} + (\alpha - 1) \sum_{i=1}^{n} \frac{2x_i^5}{(1 + \lambda x_i + \beta x_i^2)^3}$$

$$- \sum_{i=1}^{n} \alpha(\alpha - 1)(\alpha - 2)x_i^5(1 + \lambda x_i + \beta x_i^2)^{\alpha-3},$$

$$L_{333} = \sum_{i=1}^{n} \frac{16\, x_i^3}{(\lambda + 2\beta x_i)^3} + (\alpha - 1) \sum_{i=1}^{n} \frac{2x_i^6}{(1 + \lambda x_i + \beta x_i^2)^3}$$

$$- \sum_{i=1}^{n} \alpha(\alpha - 1)(\alpha - 2)x_i^6(1 + \lambda x_i + \beta x_i^2)^{\alpha-3}.$$

After substitution, the Eqs. (32-36) reduces like Lindleys integral, therefore, for the Bayes estimates of the parameter $\alpha$,

If $u(\hat{\alpha}, \hat{\lambda}, \hat{\beta}) = \hat{\alpha}$ then

$$\hat{\alpha}_{BS} = \hat{\alpha} + \frac{a-1-b\,\hat{\alpha}}{\hat{\alpha}} \sigma_{11} + \frac{c-1-d\,\hat{\lambda}}{\hat{\lambda}} \sigma_{12} + \frac{e-1-f\,\hat{\beta}}{\hat{\beta}} \sigma_{13} + \tfrac{1}{2}[A\sigma_{11}+B\sigma_{21}+C\sigma_{31}].$$

(40)

and similarly the Bayes estimate for $\lambda$ under SELF is,

If $u(\hat{\alpha}, \hat{\lambda}, \hat{\beta}) = \hat{\lambda}$ then

$$\hat{\lambda}_{BS} = \hat{\lambda} + \frac{a-1-b\,\hat{\alpha}}{\hat{\alpha}} \sigma_{21} + \frac{c-1-d\,\hat{\lambda}}{\hat{\lambda}} \sigma_{22} + \frac{e-1-f\,\hat{\beta}}{\hat{\beta}} \sigma_{23}$$

$$+ \tfrac{1}{2}[A\sigma_{12}+B\sigma_{22}+C\sigma_{32}].$$

(41)

and similarly the Bayes estimate $\beta$ for under SELF is,

If $u(\hat{\alpha}, \hat{\lambda}, \hat{\beta}) = \hat{\beta}$ then

$$\hat{\beta}_{BS} = \hat{\beta} + \frac{a-1-b\,\hat{\alpha}}{\hat{\alpha}} \sigma_{31} + \frac{c-1-d\,\hat{\lambda}}{\hat{\lambda}} \sigma_{32} + \frac{e-1-f\,\hat{\beta}}{\hat{\beta}} \sigma_{33}$$

$$+ \tfrac{1}{2}[A\sigma_{13}+B\sigma_{23}+C\sigma_{33}].$$

(42)

Further, the Bayes estimates of the reliability function and hazard function under SELF are given by

**Reliability:**

If $u(\hat{\alpha}, \hat{\lambda}, \hat{\beta}) = e^{\left[1-(1+\hat{\lambda}t+\hat{\beta}t^2)^{\hat{\alpha}}\right]}$,

then the corresponding derivatives are

$$u_1 = -e^{\left[1-(1+\lambda t+\beta t^2)^\alpha\right]} (1 + \lambda t + \beta t^2)^\alpha \ln(1 + \lambda t + \beta t^2),$$

$$u_{11} = e^{\left[1-(1+\lambda t+\beta t^2)^\alpha\right]} (1 + \lambda t + \beta t^2)^\alpha [\ln(1 + \lambda t + \beta t^2)]^2 [(1 + \lambda t + \beta t^2)^\alpha - 1],$$

$$u_{12} = te^{\left[1-(1+\lambda t+\beta t^2)^\alpha\right]} (1 + \lambda t + \beta t^2)^{\alpha-1}[\alpha \ln(1 + \lambda t + \beta t^2)((1 + \lambda t + \beta t^2)^\alpha - 1) - 1],$$

$$u_{13} = t^2 e^{\left[1-(1+\lambda t+\beta t^2)^\alpha\right]} (1 + \lambda t + \beta t^2)^{\alpha-1}[\alpha \ln(1 + \lambda t + \beta t^2)((1 + \lambda t + \beta t^2)^\alpha - 1) - 1],$$

$$u_2 = -\alpha t e^{\left[1-(1+\lambda t+\beta t^2)^\alpha\right]} (1 + \lambda t + \beta t^2)^{\alpha-1},$$

$$u_{22} = e^{\left[1-(1+\lambda t+\beta t^2)^\alpha\right]} [(\alpha t(1 + \lambda t + \beta t^2)^{\alpha-1})^2 - \alpha(\alpha - 1)t^2(1 + \lambda t + \beta t^2)^{\alpha-2}],$$

$$u_{23} = t^3 e^{\left[1-(1+\lambda t+\beta t^2)^\alpha\right]} (1 + \lambda t + \beta t^2)^{\alpha-2}[\alpha(1 - \alpha) + \alpha^2(1 + \lambda t + \beta t^2)^\alpha],$$

$$u_3 = -\alpha t^2 e^{\left[1-(1+\lambda t+\beta t^2)^\alpha\right]} (1 + \lambda t + \beta t^2)^{\alpha-1},$$

$$u_{33} = e^{\left[1-(1+\lambda t+\beta t^2)^\alpha\right]} [(\alpha t^2(1 + \lambda t + \beta t^2)^{\alpha-1})^2 - \alpha(\alpha - 1)t^4(1 + \lambda t + \beta t^2)^{\alpha-2}],$$

remaining L and $(a_1, a_2, a_3, a_4, a_5)$ terms are same as above. Therefore, reliability estimate is;

$$\hat{R}_{BS}(t) = e^{\left[1-(1+\hat{\lambda}t+\hat{\beta}t^2)^{\hat{\alpha}}\right]} + (u_1 a_1 + u_2 a_2 + u_3 a_3 + a_4 + a_5)$$
$$+ \frac{1}{2}[A(u_1\sigma_{11}+u_2\sigma_{12}+u_3\sigma_{13})+B(u_1\sigma_{21}+u_2\sigma_{22}+u_3\sigma_{23})+C(u_1\sigma_{31}+u_2\sigma_{32}+u_3\sigma_{33})].$$

(43)

**Hazard:** In the case of hazard function,

If $u(\hat{\alpha}, \hat{\lambda}, \hat{\beta}) = \hat{\alpha}(\hat{\lambda} + 2\hat{\beta}t)(1 + \hat{\lambda}t + \hat{\beta}t^2)^{\hat{\alpha}-1}$,

then the corresponding derivatives are

$$u_1 = (\lambda + 2\beta t)(1 + \lambda t + \beta t^2)^{\alpha-1}[1 + \alpha \ln(1 + \lambda t + \beta t^2)],$$

$$u_{11} = (\lambda + 2\beta t)(1 + \lambda t + \beta t^2)^{\alpha-1} \ln(1 + \lambda t + \beta t^2)[2 + \alpha \ln(1 + \lambda t + \beta t^2)],$$

$$u_{12} = (1 + \lambda t + \beta t^2)^{\alpha-1}[1 + \alpha \ln(1 + \lambda t + \beta t^2)] + t(\lambda + 2\beta t)(1 + \lambda t + \beta t^2)^{\alpha-2}[(2\alpha - 1) + \alpha(\alpha - 1) \ln(1 + \lambda t + \beta t^2)],$$

$$u_{13} = 2t(1 + \lambda t + \beta t^2)^{\alpha-1}[1 + \alpha \ln(1 + \lambda t + \beta t^2)] + t^2(\lambda + 2\beta t)(1 + \lambda t + \beta t^2)^{\alpha-2}[(2\alpha - 1) + \alpha(\alpha - 1) \ln(1 + \lambda t + \beta t^2)],$$

$$u_2 = \alpha(1 + \lambda t + \beta t^2)^{\alpha-1} + \alpha(\alpha - 1)t(\lambda + 2\beta t)(1 + \lambda t + \beta t^2)^{\alpha-2},$$

$$u_{22} = \alpha(\alpha - 1)[2t(1 + \lambda t + \beta t^2)^{\alpha-2} + (\alpha - 2)t^2(\lambda + 2\beta t)(1 + \lambda t + \beta t^2)^{\alpha-3}],$$

$$u_{23} = \alpha(\alpha - 1)[3t^2(1 + \lambda t + \beta t^2)^{\alpha-2} + (\alpha - 2)t^3(\lambda + 2\beta t)(1 + \lambda t + \beta t^2)^{\alpha-3}],$$

$$u_3 = 2\alpha t(1 + \lambda t + \beta t^2)^{\alpha-1} + \alpha(\alpha - 1)t^2(\lambda + 2\beta t)(1 + \lambda t + \beta t^2)^{\alpha-2},$$

$$u_{33} = \alpha(\alpha - 1)[4t^3(1 + \lambda t + \beta t^2)^{\alpha-2} + (\alpha - 2)t^4(\lambda + 2\beta t)(1 + \lambda t + \beta t^2)^{\alpha-3}],$$

remaining L and $(a_1, a_2, a_3, a_4, a_5)$ terms are same as above. Therefore, reliability estimate is;

$$\hat{h}_{BS}(t) = \hat{\alpha}(\hat{\lambda} + 2\hat{\beta}t)(1 + \hat{\lambda}t + \hat{\beta}t^2)^{\hat{\alpha}-1} + (u_1a_1 + u_2a_2 + u_3a_3 + a_4 + a_5)$$
$$+ \frac{1}{2}[A(u_1\sigma_{11} + u_2\sigma_{12} + u_3\sigma_{13}) + B(u_1\sigma_{21} + u_2\sigma_{22} + u_3\sigma_{23}) + C(u_1\sigma_{31} + u_2\sigma_{32} + u_3\sigma_{33})].$$

(44)

## 5.2 Markov Chain Monte Carlo Method

In this subsection, we discuss about Gibbs sampling procedure to generate sample from posterior distribution. For more details about Markov Chain Monte Carlo Method (MCMC) see **Smith** and **Roberts** (1993), **Hastings** (1970) and **Singh et al.** (2013). **Chen** and **Shao** (2000) developed a Monte Carlo method for using importance sampling to compute HPD (highest probability density) intervals for the parameters of interest or any function of them. Thus utilizing the concept of Metropolis Hastings (M-H) under Gibbs sampling procedure generate sample from the posterior density function (31) under the assumption that parameters $\alpha, \lambda$ and $\beta$ have independent gamma density function with hyper have independent gamma density function with hyper parameters (a, b), (c, d) and (g, f) respectively. To implement this technique we consider full conditional posterior densities of $\alpha, \lambda$ and $\beta$ as; parameters (a, b), (c, d) and (g, f) respectively. To implement this technique we consider full conditional posterior densities of $\alpha, \lambda$ and $\beta$ as;

$$\pi(\alpha|\lambda,\beta,X) \propto \alpha^{n+a-1} e^{-b\alpha} e^{\sum_{i=1}^{n}\left(1 - (1+\lambda x_i + \beta x_i^2)^\alpha\right)} \prod_{i=1}^{n} \left(1 + \frac{2\beta}{\lambda}x_i\right)(1 + \lambda x_i + \beta x_i^2)^{\alpha-1},$$

(45)

$$\pi(\lambda|\alpha,\beta,X) \propto \lambda^{n+c-1} e^{-d\lambda} e^{\sum_{i=1}^{n}\left(1 - (1+\lambda x_i + \beta x_i^2)^\alpha\right)} \prod_{i=1}^{n} \left(1 + \frac{2\beta}{\lambda}x_i\right)(1 + \lambda x_i + \beta x_i^2)^{\alpha-1},$$

(46)

$$\pi(\beta|\alpha,\lambda,X) \propto \beta^{g-1} e^{-f\beta} e^{\sum_{i=1}^{n}\left(1 - (1+\lambda x_i + \beta x_i^2)^\alpha\right)} \prod_{i=1}^{n} \left(1 + \frac{2\beta}{\lambda}x_i\right)(1 + \lambda x_i + \beta x_i^2)^{\alpha-1}.$$

(47)

**M-H** under Gibbs sampling algorithm consist the following steps:

Step 1: Generate $\alpha, \lambda$ and $\beta$ from (45), (46) and (47) respectively.

Step 2: Obtain the posterior sample $(\alpha_1, \lambda_1, \beta_1), (\alpha_2, \lambda_2, \beta_2), \ldots, (\alpha_M, \lambda_M, \beta_M)$ by repeating step 1, M times.

Step 3: The Bayes estimates of the parameters i.e. $\alpha, \lambda, \beta$, Reliability function $R(t)$ and Hazard function $h(t)$ with respect to the SELF are given as;

$$\hat{\alpha}_S^{MC} = [E_\pi(\alpha \backslash X)] \approx \left(\frac{1}{M} \sum_{k=1}^{M} \alpha_k\right), \tag{48}$$

$$\hat{\lambda}_S^{MC} = [E_\pi(\lambda \backslash X)] \approx \left(\frac{1}{M} \sum_{k=1}^{M} \lambda_k\right), \tag{49}$$

$$\hat{\beta}_S^{MC} = [E_\pi(\beta \backslash X)] \approx \left(\frac{1}{M} \sum_{k=1}^{M} \beta_k\right), \tag{50}$$

$$\hat{R}(t)_S^{MC} = [E_\pi(R(t) \backslash \alpha, \lambda, \beta, X)] \approx \left(\frac{1}{M} \sum_{k=1}^{M} e^{\left[1-(1+\lambda_k t+\beta_k t^2)^{\alpha_k}\right]}\right) \tag{51}$$

and

$$\hat{h}(t)_S^{MC} = [E_\pi(h(t) \backslash \alpha, \lambda, \beta, X)] \approx \left(\frac{1}{M} \sum_{k=1}^{M} \alpha_k(\lambda_k + 2\beta_k t)(1 + \lambda_k t + \beta_k t^2)^{\alpha_k - 1}\right) \tag{52}$$

respectively.

Step 4: After extracting the posterior samples we can easily construct the 95% HPD credible intervals for $\alpha, \lambda$ and $\beta$. Therefore for this purpose order $\alpha_1, \alpha_2, \ldots, \alpha_N$ as $\alpha_{(1)} < \alpha_{(2)} < \cdots < \alpha_{(N)}, \lambda_1, \lambda_2, \ldots, \lambda_N$ as $\lambda_{(1)} < \lambda_{(2)} < \cdots < \lambda_{(N)}$ and $\beta_1, \beta_2, \ldots, \beta_N$ as $\beta_{(1)} < \beta_{(2)} < \cdots < \beta_{(N)}$. Then $100(1-\vartheta)\%$ credible intervals of $\alpha, \lambda$ and $\beta$ are $\left((\alpha_{(1)}, \alpha_{[N(1-\vartheta)+1]}), \ldots, (\alpha_{[N\vartheta]}, \alpha_{(N)})\right), \left((\lambda_{(1)}, \lambda_{[N(1-\vartheta)+1]}), \ldots, (\lambda_{[N\vartheta]}, \lambda_{(N)})\right)$ and $\left((\beta_{(1)}, \beta_{[N(1-\vartheta)+1]}), \ldots, (\beta_{[N\vartheta]}, \beta_{(N)})\right)$.

Here $[x]$ denotes the greatest integer less than or equal to $X$. Then the HPD credible interval which has the shortest length.

## 6. Real Data Analysis

In this section, we study a real data set to illustrate how the proposed methodology can be applied in real life phenomenon. To check the validity of proposed model, Akaike information criterion (AIC) and Bayesian information criterion (BIC) have been discussed see Table 1. Further, we have also provided empirical cumulative distribution function (ECDF) plot and theoretical cumulative distribution function (CDF) plots for maximum likelihood estimator (MLE) as well as Bayes estimator of the parameters see figure of ECDF. After all, it is observed that proposed model works quite well. The considered data are the failure times of the air conditioning system of an air-plane taken from of size *n= 30* see **Linhart** and **Zucchini** (1986).

In this case we have fitted the four distributions namely exponential, exponentiated exponential, gamma and Weibull. Both estimation procedures have been taken into account for the considered real data set. The considered methodology can be illustrated as follows;

$AIC = -2 \ln L(X, \theta) - 2k$

$BIC = -2 \ln L(X, \theta) - k \ln(n)$

where, $L(X, \theta)$ is the likelihood function, $k$ is the number of parameters associated with model.

*Table 1*: Table shows the values of various adaptive measures for different models regarding fitting of the considered real data

| Model | $-\log L$ | AIC | BIC |
|---|---|---|---|
| $ED(\theta)$ | 152.629 | 307.259 | 308.661 |
| $EED(\alpha, \beta)$ | 152.205 | 308.411 | 311.213 |
| $Gamma(\alpha, \beta)$ | 152.167 | 308.334 | 311.137 |
| $Weibull(\alpha, \beta)$ | 151.949 | 307.878 | 310.681 |
| $ExED(\alpha, \beta)$ | 151.582 | 307.163 | 309.965 |
| $MExED(\alpha, \lambda, \beta)$ | 151.349 | 296.698 | 292.494 |

In classical set-up the maximum likelihood estimates (MLEs) of $\alpha, \lambda, \beta$, reliability function and hazard function ($R$(t), $h$(t)) are calculated as (0.22, 0.048, 0.01), (8.086×10$^{-14}$, 0.572) respectively. The 95% asymptotic confidence intervals of $\alpha, \lambda$ and $\beta$ based on fisher information matrix are obtained as (0, 75.24), (0, 32.005) and (0, 392.695) respectively.

## 7. Conclusion

This paper introduces a new model positive data. The scale-exponential distribution can be seen as a particular case of the new model. It is shown that the distribution function, hazard function and moment function can be obtained in closed form. We have considered the classical and Bayesian estimation of unknown parameters and reliability characteristics in modified extension of exponential distribution. From the simulation we can obtains that the Bayes estimates with non-informative prior behave like the maximum likelihood estimates, but for informative prior, the Bayes estimates behave much better than the maximum likelihood estimates.